# Role of Phonon Scattering in Graphene Nanoribbon Transistors: Non-Equilibrium Green's Function Method with Real Space Approach

Youngki Yoon[†], Dmitri E. Nikonov[‡], and Sayeef Salahuddin[†a)]

[†]Department of Electrical Engineering and Computer Sciences, University of California, Berkeley, CA 94720

[‡]Technology and Manufacturing Group, Intel Corp., SC1-05, Santa Clara, CA 95052

**ABSTRACT**

Mode space approach has been used so far in NEGF to treat phonon scattering for computational efficiency. Here we perform a more rigorous quantum transport simulation in real space to consider interband scatterings as well. We show a seamless transition from ballistic to dissipative transport in graphene nanoribbon transistors by varying channel length. We find acoustic phonon (AP) scattering to be the dominant scattering mechanism within the relevant range of voltage bias. Optical phonon scattering is significant only when a large gate voltage is applied. In a longer channel device, the contribution of AP scattering to the dc current becomes more significant.

[a)]E-mail: sayeef@eecs.berkeley.edu

Since the experimental demonstration in 2004 of the world's first two-dimensional (2-D) material – graphene,[1] there has been significant interest in planar carbon structures.[2] Although a 2-D graphene is a semi-metal and hence always electrically conductive, what makes it attractive for electronic applications is the fact that its bandgap can be opened up by imposing quantum confinement in one direction.[3] The quasi-one-dimensional strip of carbon, so-called graphene nanoribbon (GNR), has been intensively explored for the potential applications of field-effect transistors, resulting in promising experimental demonstrations.[4-6] In order to understand the unique transport mechanism in GNR, self-consistent atomistic quantum simulations have been performed within ballistic approximation.[7-9] In practice, however, long channel GNR field-effect transistors (FETs) show dissipative transport,[10] which indicates that scattering mechanisms should be considered in realistic device simulations. Interestingly however, electron-phonon scattering cannot be overlooked even for a relatively short channel GNRFET, as will be shown in this work.

Dissipative transport simulations of GNRFETs as well as carbon nanotube (CNT) FETs have been performed so far by means of mode space approach,[11-16] which is computationally favorable since only a few relevant subbands are considered. However, mode space approach, which is effective for intraband scattering [Fig. 1(a)], has a limitation in taking interband scattering [Fig. 1(b)] into account since each subband is treated independently of another. Since narrow GNRs are more difficult to fabricate compared to small-diameter CNTs, it is likely that a large number of subbands would be relevant to the transport in GNRFETs. Even for a 2 nm-wide GNR ($N_a$ = 16 armchair-edge GNR as used in this study), the energy difference between the first and the second lowest subband is only 160 meV. Note that the coupling between the subbands caused by phonon scattering would become more significant and complicated as GNR width increases. Therefore, in this study, we use real space approach to treat all possible couplings between the subbands.

Device characteristics are calculated from the self-consistent solution of the three-dimensional Poisson equation and the Non-Equilibrium Green's Function (NEGF) equations with tight-binding approximation in the $p_z$ orbital basis set.[17] The model device structure has $N_a$ = 16 armchair-edge GNR ($E_g$ = 0.7 eV) for the channel material. N-doped source and drain are 30 nm long, respectively, with a doping density of 0.5 dopants per nm or equivalently $2.6 \times 10^{13}$ /cm$^2$. Channel length is varied from 30 to 240 nm. Double-gate geometry is used with 2.5 nm thick HfO$_2$ ($\kappa$ = 16) dielectric material. Power supply voltage of $V_{DD}$ = 0.3 V and $T$ = 300 K are assumed.

In order to model phonon scattering, in- and out-scattering self-energies are introduced as follows

$$\Sigma_S^{in/out}(E) = D_\omega \{N_\omega(E)+1\} G^{n/p}(E \pm \hbar\omega) + D_\omega N_\omega(E) G^{n/p}(E \mp \hbar\omega), \qquad (1)$$

where $D_\omega$ is electron-phonon coupling constant, $N_\omega$ is phonon occupation number in thermal equilibrium, and $G^{n(p)}$ is electron (hole) correlation function.[17] In this study, $D_\omega$ is calculated from electron-phonon interaction Hamiltonian by following the treatment of Refs. 18-20. This approach renders elastic scattering with $D_{\omega,AP}$ = 0.01 eV$^2$ and inelastic scattering with a phonon energy $\hbar\omega$ = 196meV and $D_{\omega,OP}$ = 0.07 eV$^2$. In the presence of phonon scattering, the total self-energy consists of contact self-energy and scattering self-energy, which adds additional complexity in a full self-consistent simulation. Unlike a ballistic transport, two iterative loops are needed for the solution of the transport equation since Green's function and scattering self-energy are affected by each other. One iterative loop is required for the treatment of elastic scattering and the other is needed for inelastic scattering.

In general, phonon scattering has a negative impact on the dc current of a transistor since a portion of scattered carriers travel back to the source and cannot reach the drain. First, we plotted basic device characteristics under ballistic and dissipative transport for GNRFETs with a fixed channel length, $L_{ch}$ = 30 nm (Fig. 2). Figures 2(a) and 2(b) show that acoustic phonon (AP) scattering is a dominant scattering mechanism within the normal bias range if we define $V_{off}$ = 0.2 V and $V_{on}$ = 0.5 V with $V_{DD}$ = 0.3 V. From both plots, we observe that current is decreased by roughly 14% in the presence of AP scattering. Optical phonon (OP) scattering adds only a negligible difference in current compared to AP scattering up to $V_G$ = 0.5 V. It is significant only when a higher gate voltage is applied.[14] At $V_G$ = 0.75 V, the contribution of OP to the current becomes comparable to that of AP, for the case of $L_{ch}$ = 30 nm.

Bias-dependent phonon contribution can be understood by plotting current flow. Figures 3(a) and 3(b) show energy-resolved current spectrum along the device at $V_G$ = 0.4 and 0.6 V, respectively. OP scattering is suppressed at $V_G$ = 0.4 V since the energy window for current flow in the channel is smaller than the phonon energy [Fig. 3(a)]. However, at $V_G$ = 0.6 V, a significant number of injected carriers have empty states available for scattering, and an OP can be emitted [solid arrow in Fig. 3(b)]. Note that the number of available states for OP scattering will increase as a larger gate voltage is applied, and therefore the impact of OP scattering will become more significant [Fig. 2(a)]. Figures 3(c) and 3(d) show energy-resolved current at the end of channel position ($x = 60$ nm). Although both AP and OP scattering have been considered, at $V_G = 0.4$ V, current is mostly affected by AP and the scattering is only an elastic event [Fig. 3(c)]. At $V_G = 0.6$ V, however, it exhibits the signature of OP scattering and the spectrum is skewed down to the lower energy levels [Fig. 3(d)]. The role of OP is clearly illustrated in Figs. 3(e) and 3(f), which are electron distribution plots under ballistic and dissipative transport, respectively. In the presence of OP scattering, electrons can populate certain energy levels that are forbidden in case of ballistic transport. Note that OP scattering in the drain has minor impact on the total current of the device [dashed arrows in Figs. 3(a) and 3(b)]. The reason is that, once OP is emitted in the drain region, it is very unlikely that electrons can travel back to the source due to the insufficient energy to overcome the potential barrier.

In Figs. 3(g) and 3(h), we have also shown energy-resolved current spectrum for a longer channel device ($L_{ch} = 240$ nm) at $V_G = 0.2$ and 0.6 V, respectively, where we can see a more pronounced evidence of dissipative transport. The most prominent feature is a slanted potential profile at a high gate voltage [Fig. 3(h)], which is analogous to what one usually encounters in conventional metal-oxide-semiconductor (MOS) FETs. This indicates that our atomistic quantum transport simulation can accommodate seamless transition from ballistic to quasi-ballistic or dissipative transport as the dimension of the device increases. The slanted potential profile is due to the pile-up of electrons resulting from phonon scattering in the channel region. In most cases of OP scattering, phonon emission is a relevant mechanism to the transport, but at a low gate voltage, electrons that absorb phonons can contribute to the total current of device, too [arrow in Fig. 3(g)].

Next, we have explored the impact of phonon scattering on different size of devices by plotting ballisticity, which is defined as the ratio of current in the presence of phonon scattering to ballistic current ($I_{sca}/I_{bal}$), as a function of channel length [Fig. 4(a)]. Our simulation shows that ballisticity is decreased with channel length, which indicates that the effect of phonon scattering is enhanced as the channel length increases. This is consistent with expectations since the number of scattering event would increase in a longer channel device. Ballisticity also depends on the gate voltage for a given channel length. To better understand this voltage dependence, we have plotted ballisticity as a function of gate voltage for $L_{ch}$ = 240 nm [Fig. 4(b)]. The results are summarized as follows: First, the effect of OP is significant only at high gate voltages ($V_G$ > 0.55 V), which is consistent with what we have observed in Fig. 2. However, in the long channel device (240 nm), the main contribution of phonon results from AP (~80%) even at $V_G$ = 0.75 V. The effect of OP scattering becomes minor (~20%) since most electrons that emit phonons in the long channel may not have sufficient energy to overcome the potential drop imposed by the slanted potential profile and finally exit to the drain [Fig. 3(h)]. Second, with AP scattering, ballisticity is gradually increased or equivalently the effect of phonon scattering is decreased when the gate bias achieves the threshold. This can be explained with (i) density-of-states (DOS) of one-dimensional system and (ii) scattering mechanism. At a low $V_G$, current flows mainly near the conduction band edge ($E_c$) where carriers experience large DOS and hence large scattering rate.[21] In addition, since the second lowest subband is beyond the range of the relevant transport, carriers with positive wave vector must travel backward after an intraband elastic scattering event in this case. In contrast, however, at a high $V_G$, a considerable fraction of carriers flow well above the $E_c$ and the scattering rate could be smaller at the relevant energy levels.[21] At an increased gate voltage, the second lowest subband is involved in the transport and a portion of scattered carriers may travel forward even after a scattering event due to interband elastic scattering, which may further increase ballisticity. Third, the ballisticity is almost flat in the subthreshold region, which implies the $\log_{10}(I_D) - V_G$ curve can be shifted parallel by AP scattering. Figure 4(c) confirms this: Threshold voltage is shifted by $\Delta V_t = 23$ mV, which could be more significant if gate efficiency becomes worse.

To summarize, we have studied the effect of phonon scattering in GNRFETs by using self-consistent atomistic quantum transport simulation based on NEGF formalism with real space approach. Our simulation results show seamless transition from ballistic to dissipative transport by varying the simulated device size. The self-consistent solution of transport and electrostatic equations yields a slanted potential profile in a long channel device, as usually observed in conventional MOSFETs. AP scattering has a major impact on the current degradation within the relevant range of voltage operation, while OP scattering is significantly suppressed unless a high gate voltage is applied. Moreover, the impact of AP is even more pronounced as the size of device is increased. Finally, we have shown that the influence of phonon scattering is monotonically increased with channel length. At a high $V_G$, the effect of AP scattering on the dc current is reduced thanks to lower DOS at the relevant energy levels and the contribution from the upper subband, while OP scattering simply affects negatively the on current of the device.

**FIGURE CAPTIONS**

Figure 1. Schematic plot of (a) intraband and (b) interband scattering. Dashed and solid arrows illustrate elastic and inelastic scattering, respectively.

Figure 2. (a) $I_D - V_G$ curve at $V_D = 0.3$ V. (b) $I_D - V_D$ plot at $V_G = 0.4, 0.5$, and $0.6$ V. Dashed lines, solid lines with crosses, and solid lines with circles are for ballistic transport, acoustic phonon (AP) scattering, and both AP and optical phonon (OP) scattering, respectively.

Figure 3. Energy-resolved current spectrum at (a) $V_G = 0.4$ V and (b) $0.6$ V in the presence of phonon scattering (both AP and OP are treated) for $L_{ch} = 30$ nm GNRFETs. Solid line shows conduction band profile. Energy-resolved current at the channel-drain interface ($x = 60$ nm) under ballistic (dashed line) and dissipative transport (solid line) at (c) $V_G = 0.4$ V and (d) $0.6$ V. Electron correlation function, $G^n$ plotted on a log scale, which shows electron distribution in energy and position, under (e) ballistic and (f) dissipative transport at $V_G = 0.6$ V. Dashed line shows conduction band profile. Energy-resolved current spectrum at (g) $V_G = 0.2$ V and (h) $0.6$ V under dissipative transport (both AP and OP are considered) for $L_{ch} = 240$ nm devices. The arrows show OP scattering.

Figure 4. (a) Ballisticity ($I_{sca}/I_{bal}$) as a function of channel length at two different gate voltages. (b) Ballisticity as a function of gate voltage with AP scattering (circles) and both AP and OP scattering (squares) for $L_{ch} = 240$ nm devices. (c) $I_D - V_G$ plot for $L_{ch} = 240$ nm devices under ballistic transport (dashed line), with AP scattering (circles), and with both AP and OP scattering (solid line).

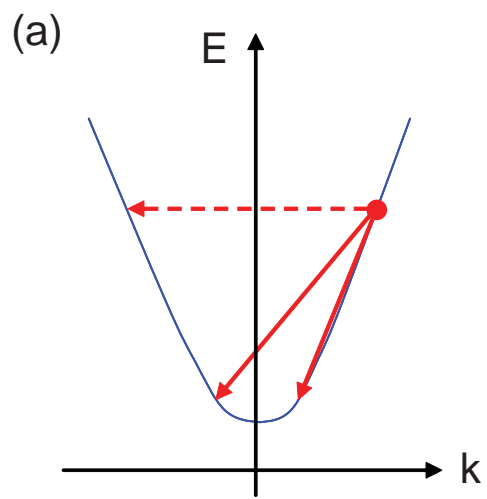 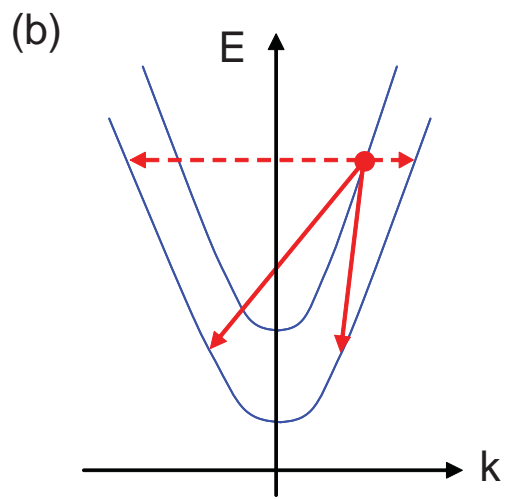

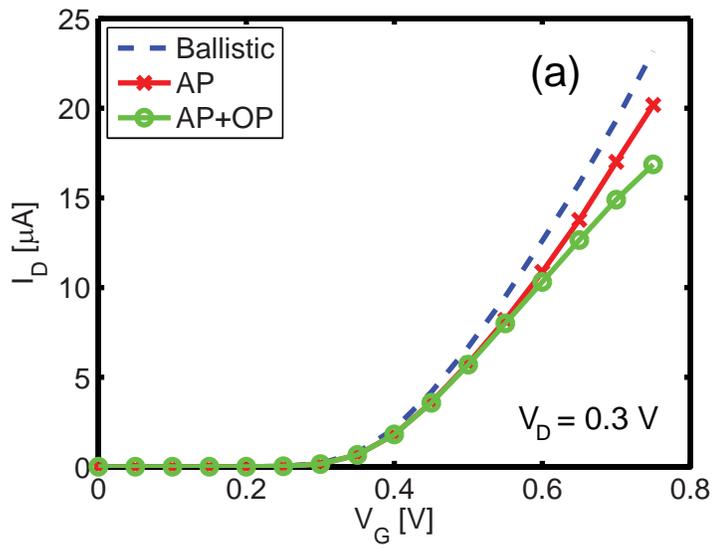 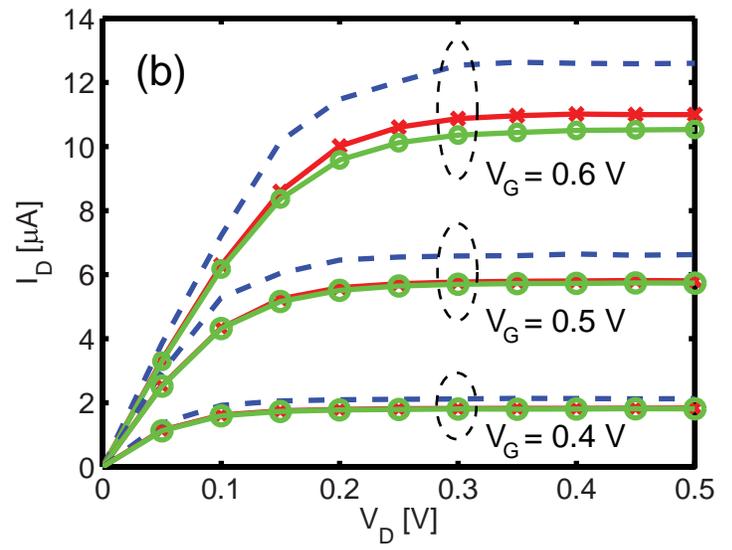

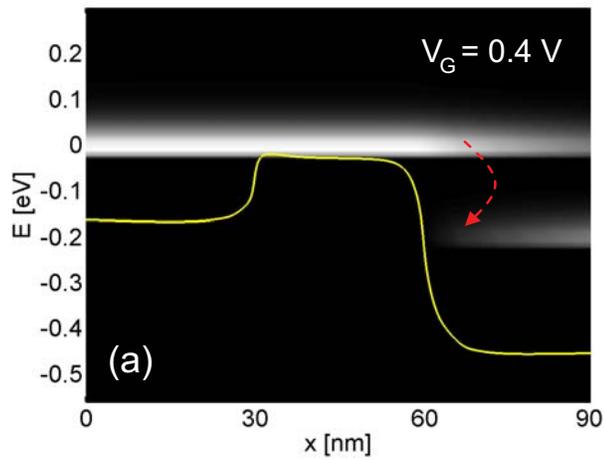
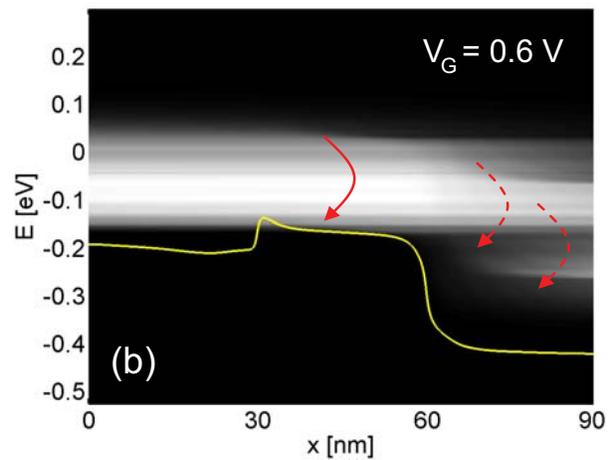
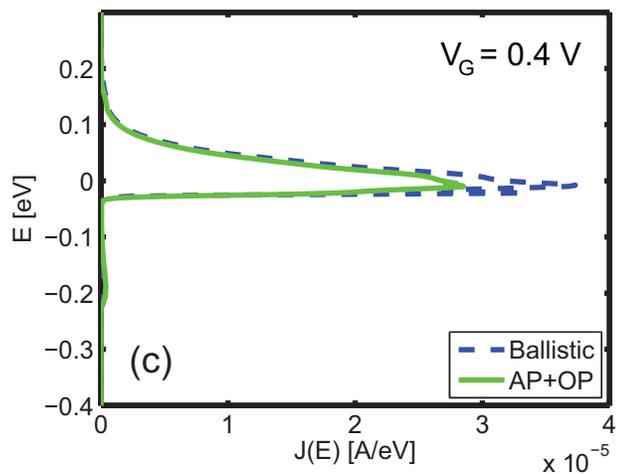
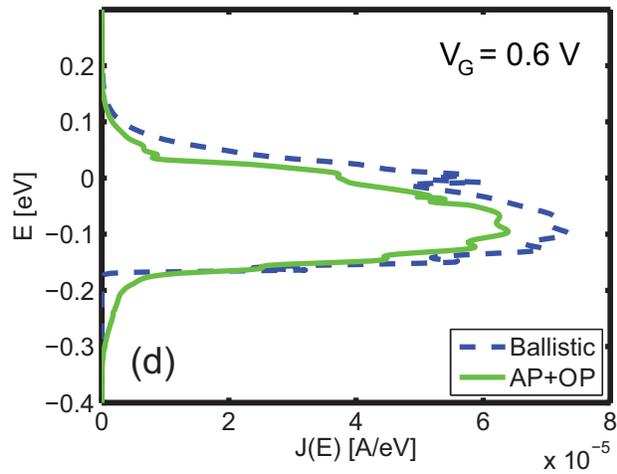
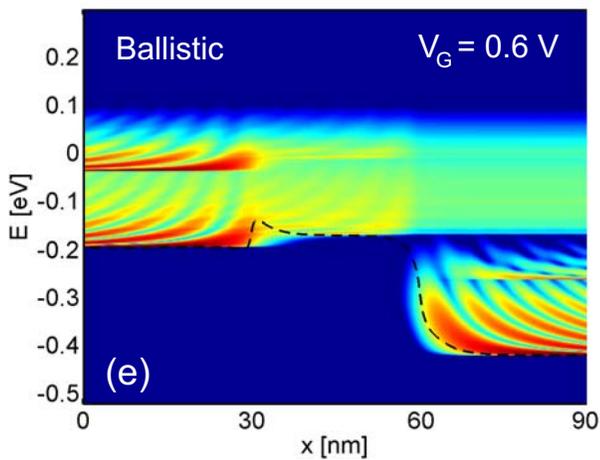
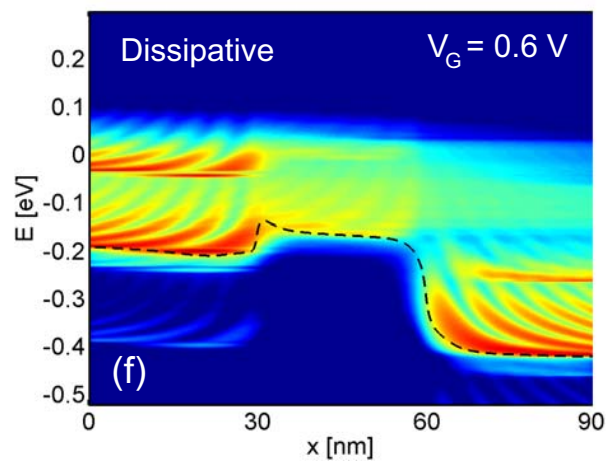
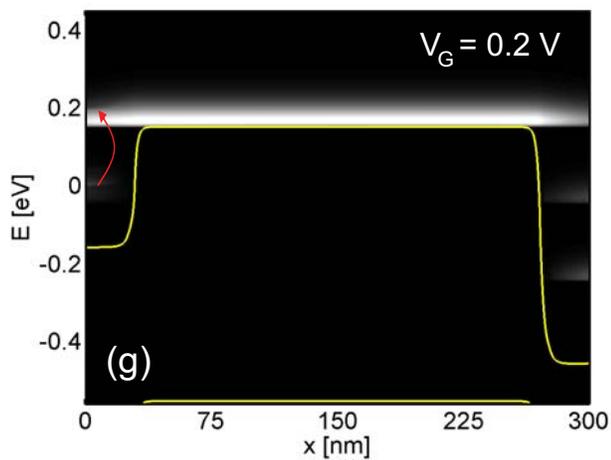
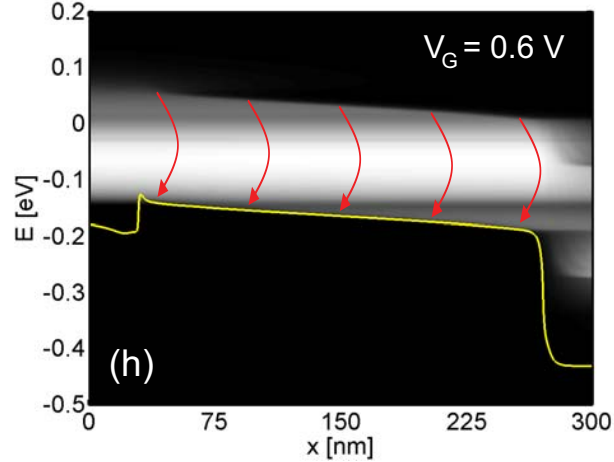

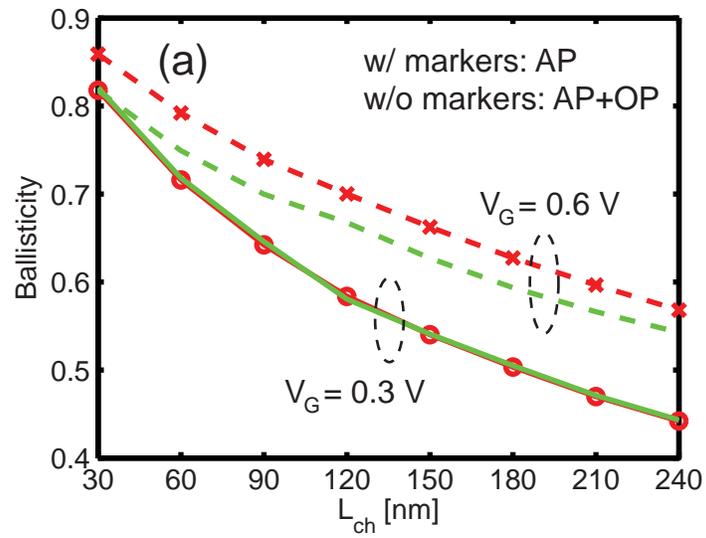

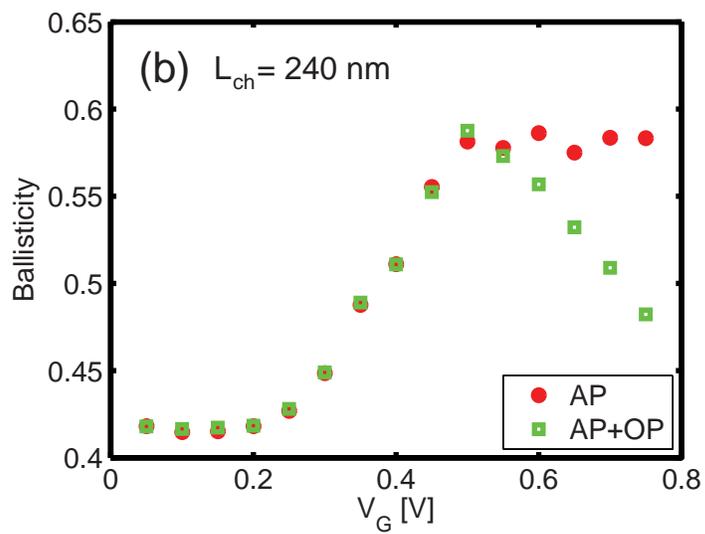

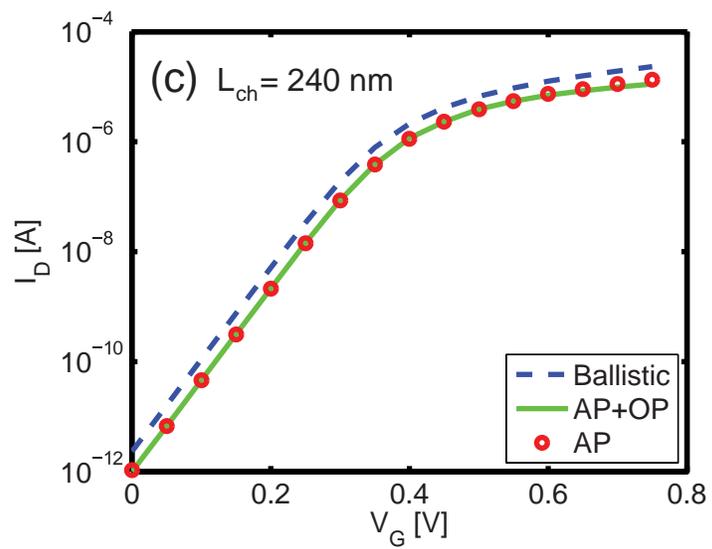